# A TEM Study of MOCVD-Grown Rutile GeO$_2$ Films

Imteaz Rahaman[1], Botong Li[1], Hunter D. Ellis[1], Brian Roy Van Devener[2], Randy C Polson[2], and Kai Fu[1, a)]

[1]*Department of Electrical and Computer Engineering, The University of Utah, Salt Lake City, UT 84112, USA*

[2]*Electron Microscopy and Surface Analysis Laboratory, The University of Utah, Salt Lake City, UT 84112, USA*

**Abstract**

Ultrawide bandgap (UWBG) semiconductors are promising for next-generation power electronics, largely attributed to their substantial bandgap and exceptional breakdown electric field. Rutile GeO$_2$ (r-GeO$_2$) emerges as a promising alternative, particularly because of its ambipolar dopability. However, research on r-GeO$_2$ is still in its infancy, and further investigation into its structural properties is essential for enhancing epilayer quality. In our previous work, we identified distinct surface morphologies—square-patterned and smooth regions—of epitaxial r-GeO$_2$ films grown on r-TiO$_2$ (001) substrates using metal-organic chemical vapor deposition (MOCVD). This research employs transmission electron microscopy (TEM) to investigate the structural characteristics of the material. The findings indicate that the square-patterned regions are crystalline, whereas the smooth regions exhibit amorphous properties. The measured lattice spacing in the (110) plane is 0.324 nm, slightly exceeding the theoretical value of 0.312 nm. This discrepancy suggests the presence of tensile strain in the r-GeO$_2$ film, resulting from lattice mismatch or thermal expansion differences with the substrate. We also observed a threading dislocation density of $1.83 \times 10^9$ cm$^{-2}$, consisting of 11.76% screw-type, 29.41% edge-type, 55.89% mixed-type dislocations, and 2.94% planar defects. These findings offer valuable insights into the growth mechanisms and defect characteristics of r-GeO$_2$.

a) Author to whom correspondence should be addressed. Electronic mail: kai.fu@utah.edu

Currently, several UWBG semiconductors, including α- and β-$Ga_2O_3$ ($E_g$=5.3-5.6 eV and $E_g$=4.48-4.9 eV, respectively)[1–4], AlGaN ($E_g$=3.4-6.0 eV)[5], and diamond ($E_g$=5.5 eV)[6], have been extensively studied for their potential in next-generation power electronics. These materials have shown promising device performance.[6–10] While these materials have shown promising device performance, they still face major challenges that limit their development and commercialization. For instance, $Ga_2O_3$ has difficulty achieving efficient p-type conductivity.[11,12] Researchers are continuously exploring new UWBG materials to address the challenges.

Recently, r-$GeO_2$ has gained attention as a promising UWBG semiconductor due to its unique physical properties and potential to overcome the challenges.[13] With a bandgap ranging from 4.44 to 4.68 eV[14–16], r-$GeO_2$ exhibits electron mobility [244 cm²/Vs ($\perp \vec{C}$) and 377 cm²/V·s (($\parallel \vec{C}$)][17], resulting in a high n-type Baliga figure of merit (27,000–35,000 ×$10^6$ $V^2$ $\Omega^{-1}$ $cm^{-2}$).[13] Additionally, its thermal conductivity is 37 $Wm^{-1}K^{-1}$ along the *a* axis and 58 $Wm^{-1}K^{-1}$ along the *c* axis, with a measurement of 51 $Wm^{-1}K^{-1}$, which is about twice that of $Ga_2O_3$.[18] Another advantage of r-$GeO_2$ is the tunability of its bandgap through alloying with other rutile-type oxides such as r-$SnO_2$, r-$TiO_2$, and r-$SiO_2$,[19,20] similar to III-V semiconductors.[21] Theoretical studies indicate that r-$GeO_2$ also supports p-type doping [hole mobility: 27 $cm^2$/Vs ($\perp \vec{C}$) and 29 ($\parallel \vec{C}$)].[15,17,22] Bulk r-$GeO_2$ can be produced using flux techniques[23,24] and chemical vapor transport,[25] methods which could lead to the availability of large-area $GeO_2$ wafers. Although these methods are still in the trial stage, there are challenges in creating affordable r-$GeO_2$ substrates with the necessary size and quality.

For epitaxial growth, this is crucial because flux techniques, which are commonly used, often lead to high impurity content in the wafers. Rutile $GeO_2$ films have been successfully grown on r-$TiO_2$ and sapphire substrates using techniques such as molecular beam epitaxy (MBE)[26], mist chemical vapor deposition (mist-CVD)[27,28], pulsed laser deposition (PLD)[29–31], and metal-organic CVD (MOCVD)[32]. In our recent work, we observed that r-$GeO_2$ films exhibit a unique surface morphology, with square-patterned faceted crystal formations.[32] Similarly, α-$GeO_2$ also displays distinct surface features, including spherulite patterns.[33] This raises a question: what are the structural differences in films with varying surface morphologies?[32]. To improve the material properties of r-$GeO_2$ films and enhance the performance of r-$GeO_2$-based devices, it is crucial to understand the nature of dislocations and the structural properties. In this study, we examined the structural properties of MOCVD-grown r-$GeO_2$ films on r-$TiO_2$ (001) substrates using TEM.

Undoped $GeO_2$ films were grown using an Agilis MOCVD system from Agnitron Technology. The r-$TiO_2$ (001) was chosen as the substrate due to its minimal lattice mismatch and low strain compatibility with $GeO_2$.[24,27] The growth process was conducted at a temperature of 925°C, with the chamber pressure maintained at 80 Torr. Tetraethylgermane (TEGe) and pure oxygen ($O_2$) were used as the main precursors, while argon (Ar) was used as both the carrier and shroud gas. The oxygen flow rate was set to 2000 SCCM, and the TEGe precursor to 160 SCCM. During the deposition, the susceptor rotated at 300 RPM to ensure uniform growth. Before the substrates were loaded into the MOCVD chamber, they were thoroughly cleaned with a piranha solution (a 3:1 mixture of sulfuric acid and hydrogen peroxide), followed by sequential cleaning with acetone, isopropanol, and deionized water. More information on the growth process, X-ray diffraction (XRD) results, and surface roughness (RMS) can be found (sample T-22) in our recent work.[32]

The r-GeO$_2$ film exhibits two distinct surface morphologies, as shown in Fig. 1(a): one with faceted crystals in the "square-patterned region" and the other with a smoother surface in the "smooth region". As noted in previous studies, the smoother regions exhibit 2D growth, while the square-patterned areas follow a 3D growth mode, aligning with the Stranski–Krastanov mechanism.[32]

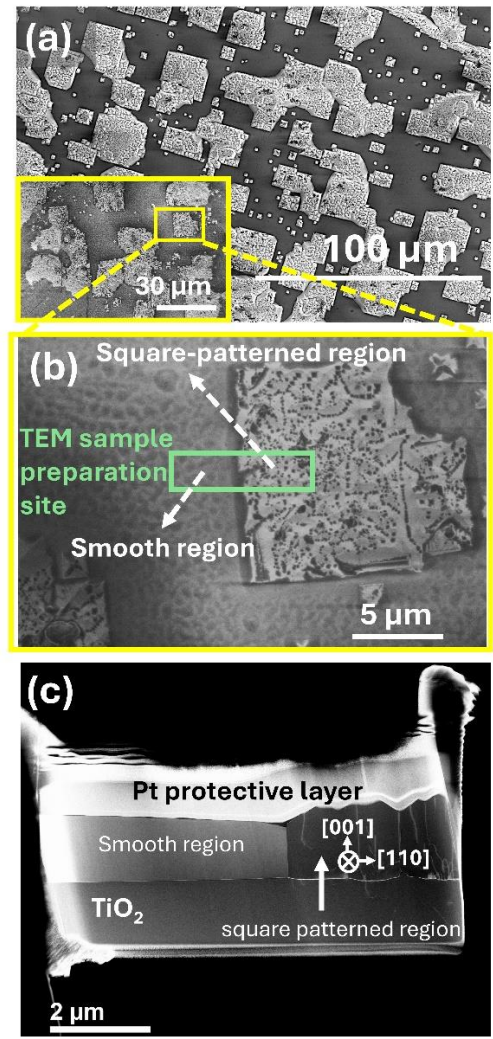

**FIG. 1.** (a) SEM images of r-GeO$_2$ films grown at 925 °C, 80 Torr, for 180 minutes (Sample T-22). The image reveals distinct regions featuring a square pattern morphology and smoother regions. (b) Higher magnification SEM of the selected area for TEM sample preparation, illustrating the transition between the square-patterned and smooth regions. (c) Cross-sectional STEM image of the lift-out lamella prepared for cross-sectional TEM imaging showing the $GeO_2$ film containing both smooth and faceted regions, Pt protective layer, and the r-$TiO_2$ (001) substrate.

To further investigate the structure, a TEM sample was prepared using a Helios Nanolab 650 Ga beam dbFIB. TEM/STEM analysis was done with a 200kV JEOL 2800 STEM equipped with dual EDS detectors. Fig. 1(b) shows the area sampled for TEM/STEM analysis, where half of the region contains the square patterned region, and the other half consists of the smooth region. Fig. 1(c) shows the cross-sectional dark field scanning transmission electron microscopy (DF-STEM) image at the r-$GeO_2$/r-$TiO_2$ interface. The r-$GeO_2$ film is clearly divided into two distinct regions, characterized by noticeable color contrast: the lighter area represents the smooth region, while the darker area corresponds to the square-patterned faceted crystal region. This contrast difference is likely due to variations in thickness or Ge content between the two regions. Additionally, the interface between the square-patterned region and the $TiO_2$ substrate is not sharp but displays a zigzag boundary.

Figure 2 shows a STEM/EDS elemental map of the r-$GeO_2$ film, comparing the smooth region (Area 1) and the square-patterned region (Area 2) on the $TiO_2$ substrate (Fig. 2(a)). Fig. 2(b) presents quantitative EDX data of r-$GeO_2$ comparing the two regions, Area 1 and Area 2. Quantitative analysis was done using an absorption-corrected Cliff-Lorimer method, using correction factors for the thickness of the sample estimated during the FIB preparation process and the bulk density of $GeO_2$, 4.2 g/cm$^3$. Absorption correction is necessary for quantitation on thick, dense materials as lower energy X-ray is more likely to be absorbed in these cases. The analysis shows that Area 1 (smooth region) has a higher oxygen (O) content (~39.79 wt%) compared to Area 2 (~29.64 wt%). Conversely, Area 2 (square patterned region) contains a higher concentration of Ge (~70.36 wt%) compared to Area 1 (~60.21 wt%). One possible explanation for the compositional differences could be the variation in $GeO_2$ film thickness between the two regions, which may affect the EDX signal, especially given the coexistence of amorphous and crystalline phases. These phases likely have different mass densities, causing variations in X-ray absorption and signal intensity. Additionally, differences in growth kinetics, such as nucleation and growth rates, could influence both the morphology and local composition of the smooth and square-patterned regions. While the measured compositional differences fall within the expected range of EDX uncertainty, further investigation using higher-precision techniques, such as TEM-EELS or XPS, would provide more definitive insights into the true stoichiometry of both regions.

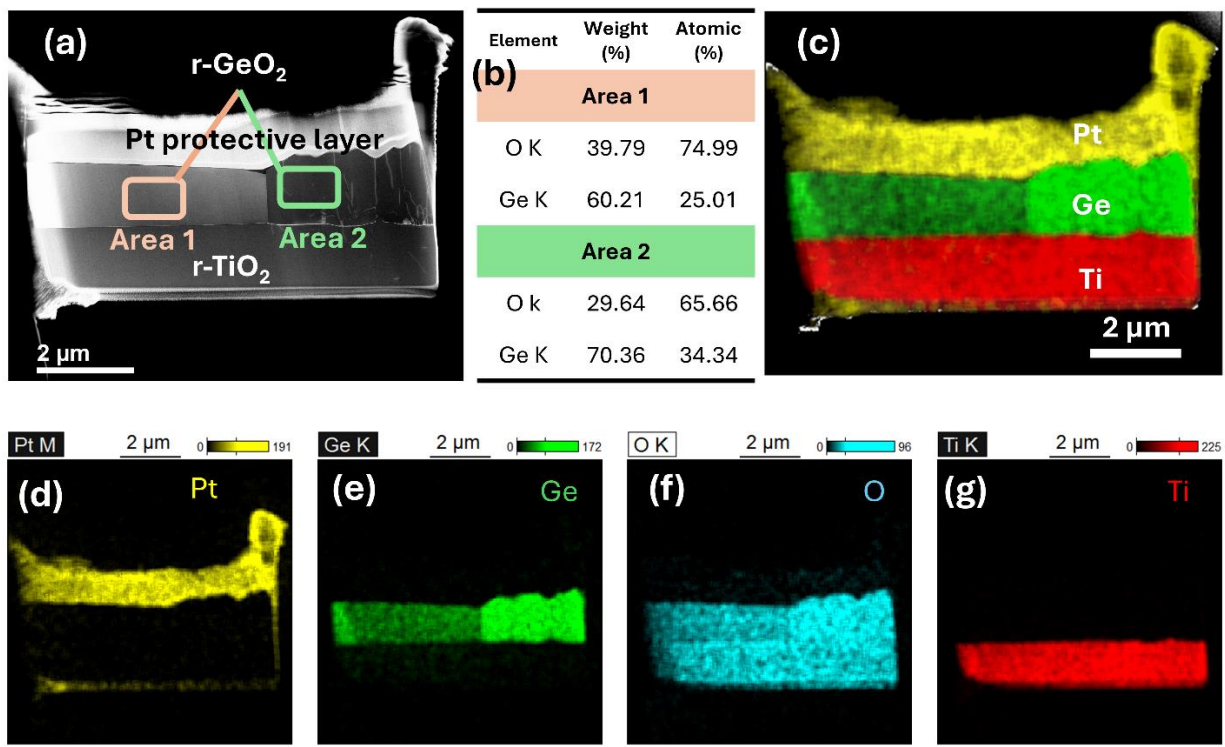

**FIG. 2.** (a) The DF-STEM image of the $r-GeO_2/r-TiO_2$ interface. (b) Quantitative EDX analysis of the two distinct areas, as indicated in (a), comparing the smooth and square-patterned regions.

(c) Corresponding background subtracted; net-count EDS elemental maps of the same region illustrate the distribution of key elements across the interface. The mapped elements are Pt (yellow), Ge (green), O (cyan), and Ti (red). (d-g) EDS elemental maps of individual elements: (d) platinum (Pt), (e) germanium (Ge), (f) oxygen (O), and (g) titanium (Ti).

Figure 2(c) shows the corresponding energy dispersive X-ray spectroscopy (EDS) elemental mapping, which provides a spatial representation of the key elements across the interface. The EDS maps were background subtracted and processed as net-counts maps. The overlay of Pt (yellow), Ge (green), O (cyan), and Ti (red) clearly delineates the r-$GeO_2$ layer from the $TiO_2$ substrate. To further clarify the chemical composition, elemental maps for each specific element are presented in Figs. 2(d-g). Pt is deposited on the surface for the FIB processing, while the Ge map (Fig. 2(e)) shows the profile of Ge in the r-$GeO_2$ epilayer. The O map (Fig. 2(f)) reveals the O profile in both r-$GeO_2$ and $TiO_2$ layers, which is consistent with the expected stoichiometry of these oxides. The Ti map (Fig. 2(g)) illustrates that Ti is restricted to the substrate region, no obvious diffusion of Ti into the r-$GeO_2$ layer.

Figure 3 presents the structural information of the r-$GeO_2$ film grown on the r-$TiO_2$ substrate, utilizing selected area electron diffraction (SAED) to distinguish between the smooth region and the square-patterned region. A bright-field scanning transmission electron microscopy (BF-STEM) image is shown in Fig. 3(a), with SAED patterns collected from four distinct regions: the smooth region (orange) and square-patterned region (green) of the r-$GeO_2$, the r-$TiO_2$ substrate (blue), and the r-$GeO_2$/r-$TiO_2$ interface (red).

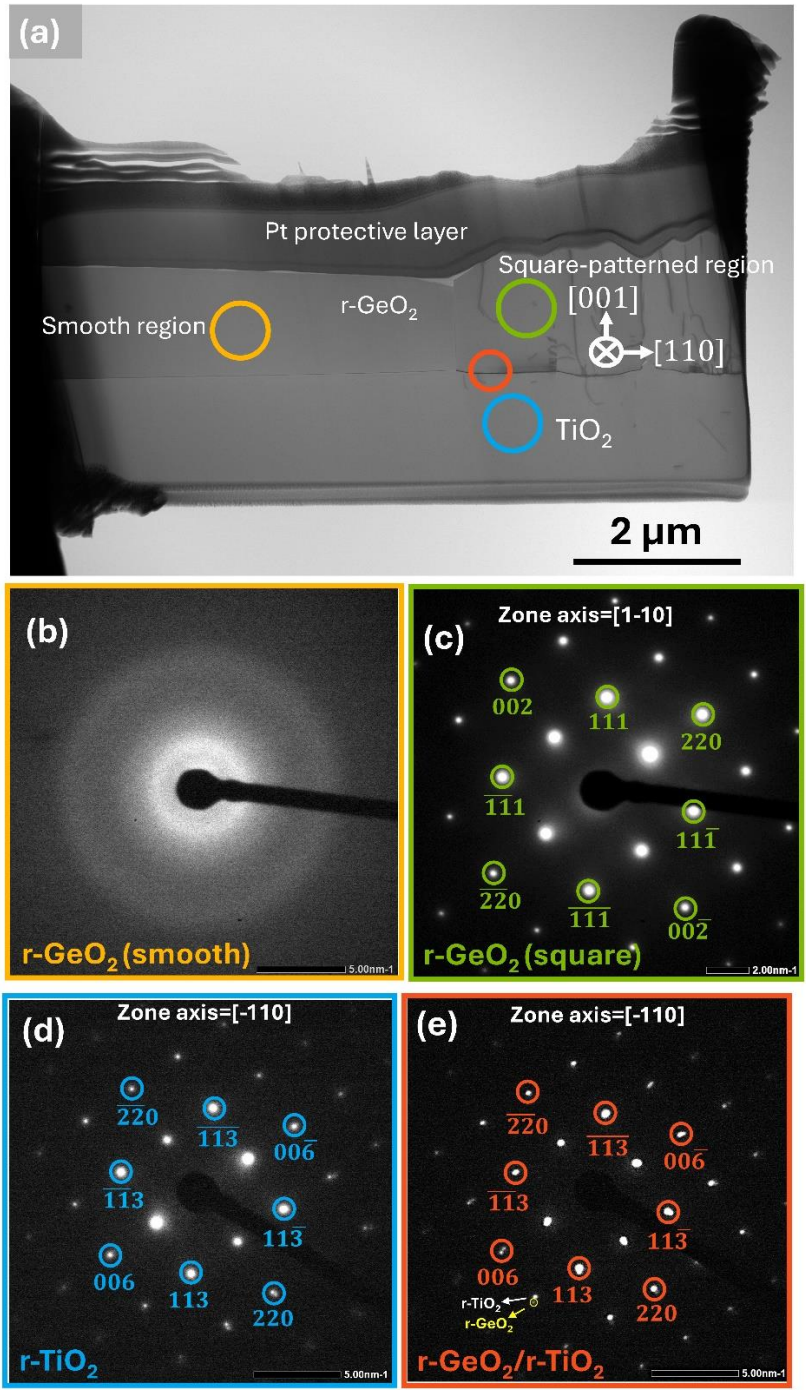

**FIG. 3.** (a) Cross-sectional BF-STEM image of the r-GeO$_2$ film on r-TiO$_2$, highlighting regions where SAED patterns were obtained: smooth (orange), square-patterned (green), r-TiO$_2$ substrate (blue), and the r-GeO$_2$/r-TiO$_2$ interface (red). (b) SAED pattern from the smooth region (orange) shows diffuse rings, indicating an amorphous structure. (c) SAED pattern from the square-patterned region (green) reveals a crystalline structure, with distinct diffraction spots corresponding to r-GeO$_2$ planes. (d) SAED pattern from the r-TiO$_2$ substrate (blue) showing clear diffraction spots of r-TiO$_2$. (e) SAED pattern from the r-GeO$_2$/r-TiO$_2$ interface (red) displaying well-defined diffraction spots from both materials.

Figure 3(b) presents the SAED pattern from the smooth region showing diffuse rings. The absence of distinct diffraction spots implies a disordered atomic structure in this region. It suggests that the smooth region is either amorphous or nanocrystalline. In contrast, Figure 3(c) shows the SAED pattern from the square patterned region, with sharp and well-defined diffraction spots corresponding to the [1-10] zone axis of r-GeO$_2$. The Miller indices for each SAED spot are also displayed, and it is determined by the crystallographic analysis software CrysTBox. These spots confirm the crystallinity of this region (green), with clearly resolved crystallographic planes such as (001), (111), and (110). Figure 3(d) displays the SAED pattern of the r-TiO$_2$, revealing distinct and well-defined diffraction spots. Figure 3(e) shows the SAED pattern at the r-GeO$_2$/r-TiO$_2$ interface, where additional diffraction spots, marked by yellow circles, are observed outside the r-TiO$_2$ pattern. These spots originate likely from r-GeO$_2$, as its lattice constants are smaller than those of r-TiO$_2$.[27,34,35] The weaker intensity of the r-GeO$_2$ spots is attributed to lattice disorder caused by the strain in the film. Moreover, the r-GeO$_2$ diffraction spots appear slightly shifted from the r-TiO$_2$ spots along the [-1-1-3] direction, while they nearly overlap along the [-1-13] direction. This shift indicates that the r-GeO$_2$ film experiences in-plane tensile stress due to lattice mismatch with the r-TiO$_2$ substrate.[27] A more detailed discussion of the strain in the *c*-axis of the r-GeO$_2$ films can be found in our recent work.[32]

Figure 4 illustrates the dislocation information of the r-GeO$_2$ film grown on the r-TiO$_2$ substrate through a combination of BF-STEM imaging and dark-field TEM using the two-beam imaging condition to highlight dislocations. Figure 4(a) shows the BF-STEM image of the r-GeO$_2$ film, viewed along the [1-10] direction. Some regions of the r-GeO$_2$ film and the underlying r-TiO$_2$ appear darker, indicating intrinsic defects in the r-GeO$_2$ or strain caused by lattice mismatch between the r-GeO$_2$ and r-TiO$_2$ substrate.[27]

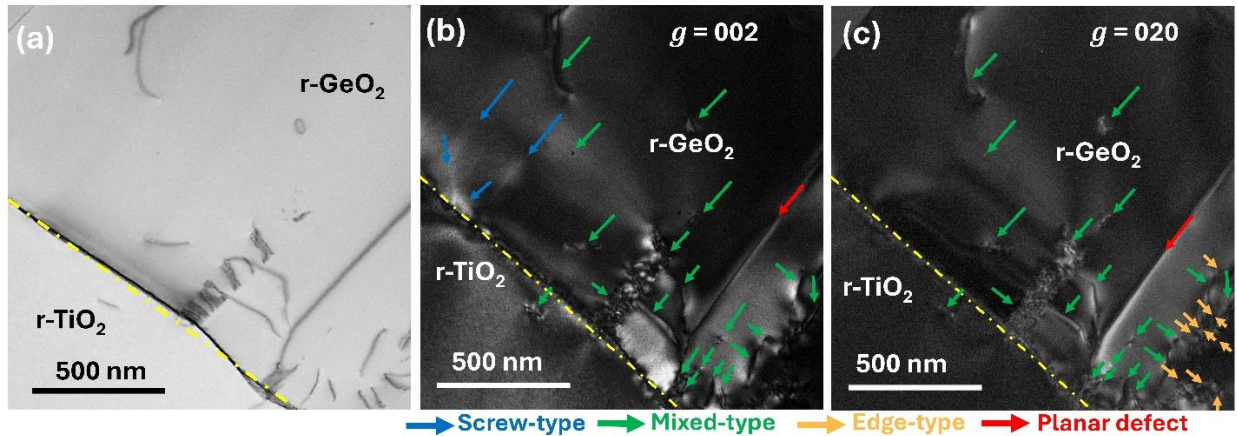

**FIG. 4.** (a) BF-STEM image of the r-GeO$_2$ film along the [1-10] direction, highlighting the surface morphology of the film and dislocation features in the square-shaped region. Cross-sectional DF-TEM images were obtained under two-beam conditions with diffraction vectors (***g***) of (b) 002 and (c) 020 along the [1-10] direction. The dislocations are categorized by different types: blue arrows mark screw-type dislocations, yellow arrows indicate edge-type dislocations, green arrows represent mixed-type dislocations, and red arrows indicate typical planar defects.

The threading dislocation density (TDD) in the film region was calculated to be $1.83 \times 10^9$ cm$^{-2}$ from Fig. 4(a) by counting the number of dislocations and dividing it by the measured area. Figures 4(b) and 4(c) show cross-sectional dark-field TEM images taken using two-beam conditions, with diffraction vectors (***g***) of 002 and 020, respectively, along the [1-10] direction. It is important to note that both Fig. 4(b) and Fig. 4(c) show the same area. Using the ***g·b*** criterion and assuming that the visible dislocations extend from the interface to the film surface, edge-type dislocations should not appear under $g$ = 002 conditions, and screw-type dislocations should not appear under $g$ = 020 conditions. This analysis allowed us to distinguish different types of dislocations, highlighted with arrows in various colors. In Fig. 4(b), the blue arrows point to screw dislocations, while the yellow arrows in Fig. 4(c) denote edge dislocations. Mixed dislocations, combining characteristics of both types, are marked with green arrows. Red arrows indicate the typical planar defects observed in the film. Screw-type, mixed-type, edge-type, and planar defects make up 11.76%, 55.89%, 29.41%, and 2.94% of the total TDD, respectively.

Figure 5 presents information on the microstructural features and crystallographic alignment of the square-patterned regions of r-GeO$_2$ film grown on the r-TiO$_2$ substrate. Figure 5(a) presents a lattice-resolution TEM image, captured along the [1-10] direction, showcasing the atomic-scale arrangement within the r-GeO$_2$ film. The alternating dark and light regions indicate thickness variations across different areas of the film, which suggest non-uniform deposition or localized differences in growth dynamics. The inset image provides a magnified view of the r-GeO$_2$ unit cell, where the lattice fringes give a measured lattice spacing, $d$ of 0.324 nm in the (110) plane, compared to a theoretical value of 0.312 nm. This slight discrepancy may be attributed to tensile strain in the r-GeO$_2$ film, likely resulting from lattice mismatch or differences in thermal expansion with the substrate. Experimental factors such as imaging resolution or minor sample tilts during TEM could also contribute to this variation. Figure 5(b) shows the inverse Fast Fourier Transform (FFT) image corresponding to the TEM image in Fig. 5(a) and the inset highlights the lattice fringes, enhancing the visibility of the crystalline structure and atomic arrangement within the r-GeO$_2$ film. It highlights the highly ordered atomic arrangement, confirming the high crystallinity of the film.

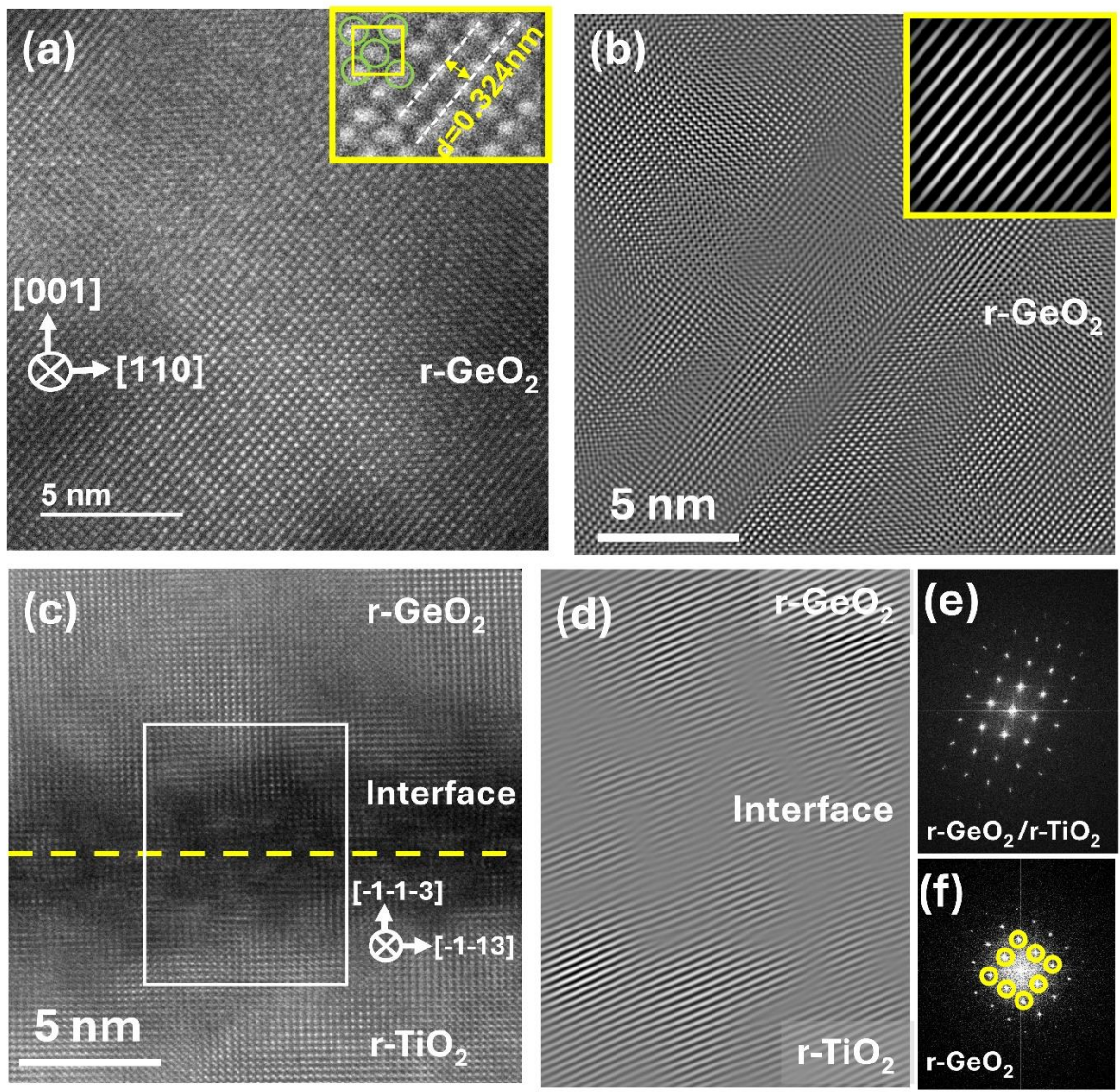

**FIG. 5.** (a) Cross-sectional TEM image of the r-GeO$_2$ film, viewed along the [1-10] direction, showcasing atomic-scale features of the film. The inset provides a magnified view of the r-GeO$_2$ unit cell, with Ge atoms represented by green circles. (b) Bragg-filtered image using the inverse Fast Fourier Transform (FFT) image corresponding to the TEM image shown in (a) with the inset showing the extracted lattice fringes. (c) Cross-sectional TEM image of the r-GeO$_2$/r-TiO$_2$ interface, viewed along the [-110] direction, displaying clear atomic-scale features and a well-defined interface, with a darker contrast between the film and the substrate. (d) Bragg-filtered

image extracted from the inverse FFT of the white box region in (c), illustrating the alignment of atomic planes at the interface. (e) FFT pattern derived from Fig. (c), confirming the epitaxial relationship between the r-GeO$_2$ film and the r-TiO$_2$ substrate. (f) FFT pattern derived from Fig. (a), indicating the crystallographic structure of the r-GeO$_2$ film.

Figure 5(c) shows the interface between r-GeO$_2$ and r-TiO$_2$, viewed along the [-110] direction with a clear contrast between the two materials. The white box highlights a region of interest for further analysis, while Fig. 5(d) focuses on the lattice fringes within this region, showing the atomic alignment at the interface. Figure 5(e) shows the FFT pattern from the r-GeO$_2$/r-TiO$_2$ interface, demonstrating well-aligned diffraction spots that confirm the crystallographic coherence between the r-GeO$_2$ film and the r-TiO$_2$ substrate. This further underscore the high-quality epitaxial growth of the r-GeO$_2$ film on the substrate, characterized by minimal defects at the interface. Figure 5(f) presents the corresponding FFT pattern of the panel (a) from the r-GeO$_2$ film, revealing sharp diffraction spots that validate the single-crystalline nature of the square-patterned region. The crystallographic orientation and the lack of misoriented domains further affirm that the MOCVD-grown r-GeO$_2$ film is epitaxially aligned with the r-TiO$_2$ (001) substrate.

In conclusion, this study investigated the structural and morphological characteristics of r-GeO$_2$ films epitaxially grown on r-TiO$_2$ (001) substrates using MOCVD. SEM image revealed two distinct regions: a square-patterned faceted crystal region and a smooth region, indicating different growth dynamics across the film. SAED patterns confirmed the crystalline nature of the square-patterned region, while the smooth region exhibited amorphous or nanocrystalline characteristics. EDS analysis indicated a uniform distribution of Ge and O within the r-GeO$_2$ layer, with no significant Ti diffusion from the substrate, ensuring a stable interface. TEM analysis using the two-beam condition revealed a TDD of $1.83 \times 10^9$ cm$^{-2}$, consisting of 2.94% planar defects, 11.76% screw-type, 29.41% edge-type, and 55.89% mixed-type dislocations. In addition, the lattice spacing in the (110) plane was measured to be 0.324 nm, slightly larger than the theoretical value of 0.312 nm, suggesting the presence of tensile strain in the r-$GeO_2$ film likely due to lattice mismatch or thermal expansion differences with the substrate. Furthermore, TEM and FFT analysis confirmed the rutile-type crystal structure of the r-$GeO_2$, reinforcing the crystallinity and epitaxial alignment of the square-patterned faceted crystal region without misoriented domains.

## AUTHOR DECLARATIONS

### Conflict of Interest

The authors have no conflicts to disclose.

### Author Contributions

Imteaz Rahaman: Data curation (lead); Formal analysis (lead); Investigation (lead); Methodology (lead), Writing-original draft (lead); Botong Li: Writing – review & editing (supporting). Hunter D. Ellis: Writing – review & editing (supporting); Brian Roy Van Devener: Data curation (equal); Writing – review & editing (supporting); Formal analysis (supporting). Randy Polson: Data curation (supporting). Kai Fu: Conceptualization (lead); Supervision (lead); Project administration (lead); Resources (lead), Writing – review & editing (lead).

## ACKNOWLEDGEMENT

The authors would like to acknowledge Miloslav Klinger for developing and providing free access to the CrysTBox software, which was used for indexing the Miller indices in the SAED patterns. Special thanks to Dr. Mike Scarpulla for his valuable feedback and brief review of this manuscript. The authors also gratefully acknowledge the support from the University of Utah start-up fund and PIVOT Energy Accelerator Grant U-7352FuEnergyAccelerator2023. Additionally,

The University of Utah provides instrumentation facilities, including the Utah Nanofab Cleanroom, The Material Characterization Meldrum, and the Nanofab Electron Microscopy and Surface Analysis Facilities.## DATA AVAILABILITY

The data that supports the findings of this study are available from the corresponding authors upon reasonable request.